\documentclass[12pt]{iopart}

\usepackage{graphicx}

\begin{document}

\title{Dark Matter \& Dark Energy from a single scalar field:
CMB spectrum and matter transfer function.}

\author{ Roberto Mainini, Silvio Bonometto}

\address{Department of Physics G.~Occhialini -- Milano--Bicocca
 University, Piazza della Scienza 3, 20126 Milano, Italy \&
I.N.F.N., Sezione di Milano}

\begin{abstract}

The dual axion model (DAM), yielding bot DM and DE form a PQ--like
scalar field solving the strong CP problem, is known to allow a fair
fit of CMB data. Recently, however, it was shown that its transfer
function exhibits significant anomalies, causing difficulties to fit
deep galaxy sample data.  Here we show how DAM can be modified to
agree with the latter data set. The modification follows the pattern
suggested to reconcile any PQ--like approach with gravity. Modified
DAM allows precise predictions which can be testable against future
CMB and/or deep sample data.

\end{abstract}

\maketitle

\section{Introduction} 
\label{sec:intro} 
A tenable cosmological model should include at least two dark
components, cold Dark Matter (DM) and Dark Energy (DE), whose nature
is still hypothetical. If DM and DE are physically unrelated, their
similar density, in today's world and just in it, is purely
accidental.

Attempts to overcome this conceptual deadlock led, first of all, to
dynamical DE \cite{DDE} (for a review see \cite{PR} and references
therein), then to considering a possible DE--DM interaction
\cite{interaction}.  These options imply new parameters, the hope
being that phenomenological limits on them guide to a gradual
understanding of the microphysics involved.

An opposite pattern is followed in the proposed {\it dual axion model}
(DAM hereafter) \cite{Mainini2004}, which widens the idea of DM made
of {\it axions}. In DAM, both DM and DE derive from a single complex
scalar field $\Phi$, being its quantized phase and modulus,
respectively. $\Phi$ assures $CP$ conservation in strong
interactions, according to Peccei \& Quinn's scheme \cite{PQ} (see
also \cite{axion,DF}), yielding dynamical DE coupled to DM.  But, once
the DE potential is selected, model parameters are fixed.

It is then quite appealing that CMB data \cite{Spergel2006} are naturally
fitted, yielding reasonable values for standard model parameters as
the primeval spectral index $n_s,$ the cosmic opacity $\tau$ and the
density parameters. On the contrary, DAM predicts rather high values
for $H_o.$

More recently \cite{pap0}, however, it was recognized that DAM leads
to a significant weakening of the Meszaros effect in the early
fluctuation evolution, causing an insufficient slope of the transfered
spectrum for $k  >\sim 0.1\, h$Mpc$^{-1}$ (see, however, \cite{Atrio} for 
a model with the opposite effect). In this paper, we show how
DAM can be improved on this point. The modified DAM predicts a
phaenomenology closer to $\Lambda$CDM, according to the value of a
suitable parameter.  As available data reasonably agree with
$\Lambda$CDM, limits on such extra parameter~can~be~set.

However, at variance from dynamical DE or coupled DE models, which
introduce one or two new parameters in top of $\Lambda$CDM's, the
whole {\it modified} DAM scheme involves the same parameter budget of
$\Lambda$CDM, with the advantage that each parameter bears a specific
physical interpretation.

Furthermore, in principle, the value of the matter density parameter
$\Omega_{om}$ fixes all parameters in the DE potential of DAM. But,
even though these parameters cannot be stringently constrained by
shortly forthcoming data, the range of the DE parameters is predicted
by the model, and this is susceptible of more immediate testing.

\section{The PQ scheme}
The strong $CP$ problem arises from the existence of multiple vacuum
states $|0_n\rangle$ in QCD: the set of the gauge transformations
${\Omega(x_\mu)}$ can be subdivided in classes ${\Omega_n(x_\mu)}$,
whose asymptotic behaviors depend on $n$ \cite{JR}. At fixed $n$, the
transformations $\Omega_n(x_\mu)$ can be distorted into each other
with continuity, while this is impossible between $\Omega_n(x_\mu)$,
with different $n$ values. Although in classical field theory no
communication between different--$n$ gauge sectors is allowed, in
quantum field theory tunneling is possible, thanks to instanton
effects. Any vacuum state is therefore a superposition
\begin{equation}
|0_\theta\rangle = \sum_n |0_n\rangle \exp(in\theta)
\end{equation}
with a suitable $\theta$ phase. The effects of varying $\theta$ can be
recast into variations of a non--perturbative term in the QCD Lagrangian
\begin{equation}
{\cal L}_\theta = {\alpha_s \over 2\pi} \theta \, G \cdot {\tilde G}~;
\label{eq:n1}
\end{equation}
here $\alpha_s$ is the strong coupling constant, $G$ and $ {\tilde G}$
are the gluon field tensor and its dual.  However, chiral
transformations also change the vacuum angle, so that, when the quark
mass matrix $\cal M$ is diagonalized, the $\theta$--parameter receives
another contribution from the EW (electro-weak) sector, becoming
\begin{equation}
\theta_{eff} = \theta + Arg ~det ~{\cal M}~.
\end{equation} 
The Lagrangian term (\ref{eq:n1}) can be reset in the form of a
4--divergence and causes no change of the equations of motion. It
however violates $CP$ and yields a neutron electric moment $d_n \simeq
5 \cdot 10^{-16} \theta_{eff} {\rm ~e~cm}$, conflicting with the
experimental limit $d_n < \sim 10^{-25} {\rm ~e~cm}$, unless
$\theta_{eff} < \sim 10^{-10}$. The point is that the two
contributions to $\theta_{eff}$ are uncorrelated, so that there is no
reason why their sum should be so small.

PQ suppress this term by imposing an additional global chiral symmetry
$U(1)_{PQ}$, spontaneously broken at a suitable scale $F_{PQ}$. The
axion field is a Goldstone boson which turns out to be suitably
coupled to the quark sector.  The details of this coupling depend on
the model and may require the introduction of an {\it ad--hoc} heavy
quark \cite{KSDZ}.  The $U(1)_{PQ}$ symmetry suffers from a chiral
anomaly, so the axion acquires a tiny mass because of non-perturbative
effects, whose size has a rapid increase around the quark-hadron
transition scale $\Lambda_{QCD}$.  The anomaly manifests itself when a
chiral $U(1)_{PQ}$ transformation is performed on the axion field,
giving rise to a lagrangian term of the same form of the one in
eq.~(\ref{eq:n1}), which provides a potential for the axion field.

As a result, $\theta$ is effectively replaced by the dynamical axion
field. Its oscillations about the potential minimum yield axions. This
mechanism works independently of the scale $F_{PQ}$. Limits on it
arise from astrophysics and cosmology, requiring that $10^{10}
GeV < \sim F_{PQ} < \sim 10^{12}GeV$; in turn, this yields an
axion mass which lays today in the interval $10^{-6} eV < \sim m_A
< \sim 10^{-3} eV$.

In most axion models, the PQ symmetry breaking occurs when a complex
scalar field $\Phi = \phi e^{i\theta}/\sqrt{2}$, falling into one of
the minima of a NG potential
\begin{equation}
V(\Phi) = \lambda [|\Phi|^2 - F_{PQ}^2]^2 ~,
\label{eq:n2}
\end{equation}
develops a vacuum expectation value $\langle \phi \rangle= F_{PQ}$.
The $CP$-violating term, arising around quark-hadron transition when
$\bar q q$ condensates break the chiral symmetry, reads
\begin{equation}
V(\theta) = \left[\sum_q \langle 0(T)| {\bar q} q |0(T) \rangle m_q \right] 
~(1 - \cos \theta)
\label{eq:n3}
\end{equation}
($\sum_q$ extends over all quarks), so that $\theta$ is no longer
arbitrary, but shall be ruled by a suitable equation of motion. The
term in square brackets, at $T \simeq 0$, approaches $m_\pi^2 f_\pi^2$
($m_\pi$ and $f_\pi$: $\pi$--meson mass and decay constant).  In this
limit, for $\theta \ll 1$ and using $A=\theta F_{PQ}$ as axion field,
eq.~(\ref{eq:n3}) reads:
\begin{equation}
V(\theta) \simeq  {1\over 2} q^2(m_q) m_\pi^2 f_\pi^2 {A^2 \over F_{PQ}^2}~;
\label{eq:n4}
\end{equation}
here $q(m_q)$ is a function of the quark masses $m_{q_i}$; in the limit of
2 light quarks ($u$ and $d$), $q = \sqrt{m_u/m_d}(1+m_u/m_d)^{-1}$.
Here below, instead of using $A,$ the axion degrees of freedom will be
described through $\theta$ itself. Eq.~(\ref{eq:n4}), however, shows
that, when $\langle \bar q q \rangle$ is no longer zero (since $T
< \sim \Lambda_{QCD}$), the axion mass decreases with temperature
approaching the constant value $m_{A}={m_\pi f_\pi q(m_q) / F_{PQ}}$
for $T \ll \Lambda_{QCD}$. Accordingly, the equation of motion, in
the small $\theta$ limit, reads
\begin{equation} 
\ddot \theta + 2{\dot a \over a} 
\dot \theta +  a^2 m_A^2  \theta = 0~,
\end{equation} 
(here $a$ is the scale factor and dots yield differentiation with
respect to conformal time, see next Section), so that the axion field
undergoes (nearly) harmonic oscillations, as soon as $m_A$ exceeds the
expansion rate; then, his mean pressure vanishes leaving axion as a
viable candidate for cold DM \cite{DF}.

This appealing scheme has been subject to various criticisms, in
connection with quantum gravity effects and Super--Symmetries (SUSY).
According to \cite{kamion}, in order to fulfill the {\it no--hair}
theorem \cite{nohair}, \cite{nohair1}, essentially stating that black
holes cannot exhibit {\it global} charges, one or more potential terms
of the form
\begin{equation}
\label{potV}
\tilde V = m_P^4 {(\Phi \Phi^*)^q \over m_P^{2q+p} }
\left( g \Phi^{p} +  g^* {\Phi^*}^p \right)
\end{equation}
should be added. Among them, terms with $p=0$ would yield just a
$\Phi$--field self--interaction. They are however needed, in
association with $p \neq 0$ terms to build potentials of the form
\begin{equation}
\label{vg}
\tilde V_n (\phi,\theta) = g_n \phi^4  (\phi/\sqrt{2}m_P)^{n}
(1-\cos \theta)~,
\end{equation}
breaking the $U_{PQ}(1)$ invariance; $g = |g| \exp(i\delta),$ in
principle, could also be complex, but $\delta \neq 0$ causes problems
discussed in \cite{kamion}, that we avoid by taking $\delta=0.$

In order to fulfill the {\it no--hair} theorem, a term of this kind
with $n \geq 1$ should exist. The physical correction could be a
function which can be expanded in a sum of terms like (\ref{vg}), with
various $n$ values, however including $n=1.$ This might help to
recover consistency between the PQ approach and gravity, without
explicitly requiring the fine tuning $g_1 <\sim 10^{-56},$ without
which {\it CP} violations reappear. The point is that the axion mass
$\sim \Lambda_{QCD}^2/F_{PQ}$ is naturally small, while gravitational
corrections must meet the same order of magnitude starting from the
Planck mass scale, and a prescription doing so in a natural way has
not been introduced yet. This problem will not be easied in the
(modified) DAM approach.

\section{The DAM scheme}
In the DAM scheme, the NG potential in eq.~(\ref{eq:n2}) is replaced
by a potential $V(\Phi)$ admitting a tracker solution \cite{DDE},
\cite{Brax}.  The field $\Phi$ is complex and $V(\Phi)$ is $U(1)$
invariant. In the modified--DAM scheme a small symmetry breacking
term, similar to eq.~(\ref{vg}), shall also be added, which will be fully
irrelevant at large $T.$

At variance from the PQ scheme, in DAM models there is no transition
to a constant value $F_{PQ}$, which is replaced by the modulus $\phi$
itself, slowly evolving over cosmological times. At a suitable early
time, $\phi$ settles on the tracker solution and, when chiral symmetry
breaks, dynamics becomes relevant also for the $\theta$ degree of
freedom, as in the PQ case. In the modified--DAM scheme, the potential
(\ref{vg}) shall also contribute to the $\theta$ dynamics, but this
will occur at much smaller energy scales.

The $\Phi$ field, therefore, besides of providing DM through its phase
$\theta$, whose dynamics solves the strong $CP$ problem, also accounts
for DE through its modulus $\phi$. Therefore, here below, the $\theta$
and $\phi$ components will be often indicated by the indeces $_c$,
$_{de}$.

In principle, this scheme holds for any DE potential admitting tracker
solutions. Here we use the SUGRA potential \cite{Brax}
\begin{equation}
\label{sugra}
V(\Phi) = {\Lambda^{\alpha+4} \over \phi^\alpha} \exp{[4\pi (\phi/m_p)^2]}
\end{equation}
found to fit available observational data, as uncoupled DE, even
slightly better than $\Lambda$CDM \cite{gerva}, without severe
restriction on the energy scale $\Lambda$ (and/or the exponent
$\alpha$).

In this cosmology, a critical stage occurs at the quark--hadron
transition, when the invariance for phase rotations in the $\Phi$
field is broken by chiral symmetry violating term:
\begin{equation}
V(\theta) =  m^2(T,\phi) \phi^2 (1-\cos\theta)
\end{equation}
($m(T,\phi)$ is the mass of the $\theta$ field which is discussed 
in section \ref{sec:lt}). The $\theta$ phase
is then driven to move about its minimum energy configuration,
starting from a generic value. The amplitude of oscillations then
gradually decreases and, when the small--$\theta$ regime is achieved,
this component will behave as CDM.

This stage can be suitably described through numerical integration,
the essential point being that it sets the initial amount of DM.  A
fair value of DM today will then arise if the modulus $\phi$ has a
suitable value at the transition and a suitable evolution of $\phi$
and $\theta$, from then to the present epoch, will then occur.

In particular, the value of $\phi$ at the transition shall exceed the
$ F_{PQ}$ energy scale by $\sim 3$ orders of magnitude, but, to yield
a significant DE amount it must increase up to $\sim m_P$ when
approaching today.

A fair evolution of both $\phi$ and $\theta$ is then achieved by
setting $\Lambda \sim 10^{10}$GeV in the SUGRA potential. $\theta$ is
then also driven to values even smaller than in the PQ case, so that
$CP$ is apparently conserved in strong interactions. Even more
significantly, $\Omega_{o,c}$ and $\Omega_{o,de}$ (DM and DE density
parameters) are let to take fair values.

In $\Lambda$CDM models, $\Omega_{o,c}$, $\Omega_{o,de}$ and
$\Omega_{o,b}$ (baryon density parameter) are free parameters.  In
uncoupled dynamical DE models, {\it e.g.} with a SUGRA potential, a
further free parameter exists, $\alpha$ or $\Lambda.$ When a constant
DM--DE coupling is added, if must be weighted by a further parameter
$\beta$; a variable coupling case needs at least a further parameter
$\epsilon,$ altogether setting that the coupling intensity $C =
(\beta/m_P) (\phi/ m_P)^\epsilon$ (see also eq.~\ref{Cop} here below).

In the DAM scheme, once the $\Lambda$ scale is assigned, fair values
of $\Omega_{o,c}$ and $\Omega_{o,de}$ naturally and unavoidably
arise. They can be modified just only by modifying the $\Lambda$
value. The parameter budget of this scheme is similar to SCDM.

\section{Modified DAM}
The need to modify this scheme, as already outlined, arises from the
damping of the stagnation or {\it Meszaros'} effect it causes. Let us
remind first what happens, in the absence of DM--DE coupling, when
fluctuations approach the horizon before radiation--matter equality.
Before entering the horizon, DM and photon--baryon fluctuations
($\delta_c$ and $\delta_{\gamma b}$) are the same. As soon as inside
the horizon, instead, $\delta_c$ and $\delta_{\gamma b}$ have
different behaviors: $\delta_{\gamma b}$ starts to fluctuate as a
sonic wave, so that $\langle \delta_{\gamma b} \rangle = 0.$ On the
contrary, CDM fluctuations do not take part in sonic waves (CDM is
non--collisional), do not ``free--stream'' (CDM particles are
non--relativistic), fail to increase significantly because of
self--gravity, as then $\Omega_c \ll 1$ and the photon--baryon fluid,
whose density parameter is $1 - \Omega_c$ (forgetting neutrinos), is
no gravity source just because $\langle \delta_{\gamma b} \rangle =
0.$ Therefore $\delta_c$ stagnate or has just a marginal growth. If we
assume that $\delta_c,$ between horizon entry and equality, roughly
grows $\propto a^{0.4}$ (outside the horizon, in a synchronous gauge,
it is then $\delta_c \propto a^2$), we obtain the basic shape of the
transfer function ${\cal T}(k)$. Its dependence on $k$ simply arises
from the varying duration of the stagnation period.

\begin{figure}[h!]
\centering
\vskip-.1truecm
\includegraphics[height=16.cm,angle=0]{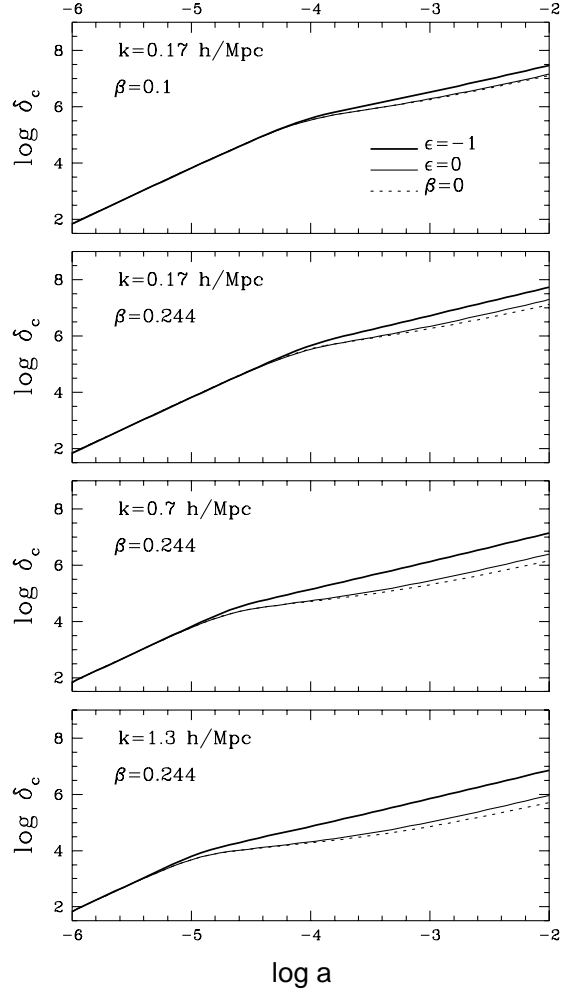}
\vskip -1.9truecm
\caption{Evolution of $\delta_c$ in the absence of coupling
($\beta=0$), for constant coupling ($\epsilon=0$) and for the DAM
model ($\beta = 0.244,~\epsilon = -1$), for different values of $k.$
For increasing $k$ values, {\it i.e.} for smaller scales, which should
undergo a longer stagnation period, the effects of (variable) coupling
become more and more significant.  }
\label{deltas}
\end{figure}

DM--DE coupling changes this scenario though two effects: it keeps
$DM$ and $DE$ densities at close values; then, interactions carried by
DE are significant and add to gravity.  The stagnation period is then
suppressed, mostly because of the enhancement of the effective
self--gravity arising from $\delta_c,$ able to beat the low $\Omega_c$
value. In \cite{pap0} the whole dynamics has been followed in detail,
suitably modifying standard linear codes. A specimen of the modified
behaviors of $\delta_c$ is given in Figure \ref{deltas}, for different
coupling intensities. The DAM case is $\beta = 0.244$ and $\epsilon =
-1.$ The case $\beta = 0.1$ and $\beta = 0$ are the cases of smaller
or vanishing coupling intensity.

To recover consistency with data, these anomalies must be prevented:
$\phi$ and $\theta$ must decouple before the relevant scales enter the
horizon. Adding a potential term
\begin{equation}
\label{-2}
\tilde V_{-2} = g \phi^2 m_P^2 (1-\cos \theta)
\end{equation} 
of the kind considered in eq.~(\ref{vg}), explicitly breaking the
$U(1)$ invariance even before the quark--hadron transition, succeeds
in doing so. The $g$ coefficient must and can be small enough, so that
this term does not perturb the whole PQ--like mechanism; it is however
possible, even for very small $g,$ that the potential (\ref{-2})
becomes significant when $\phi$ becomes large, so that $\phi$
and~$\theta$ modes eventually decouple.

We shall show that a general self consistency is then recovered for
$\Lambda$ values not far from those of the DAM, relegating the
emergence of the $V_{-2}$ term at late times. This is however enough
to let Meszaros' effect to work, so allowing a fair fit of available
data.

\section{Lagrangian theory}
\label{sec:lt}
\begin{figure}[h!]
\centering
\vskip-.2truecm
\includegraphics[height=9.cm,angle=0]{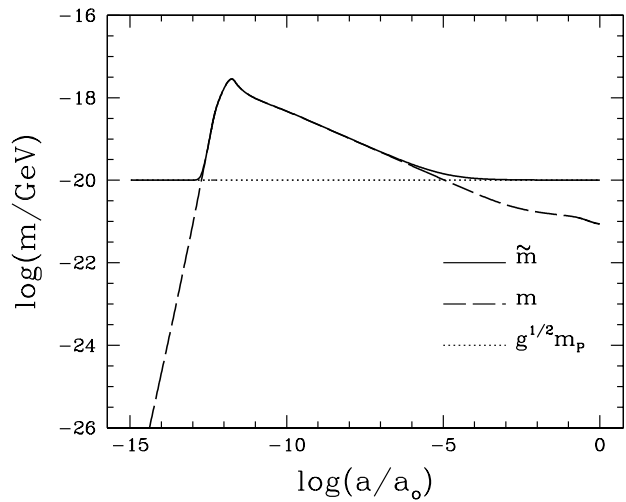}
\vskip -2.truecm
\caption{Evolution of the mass term. The break of PQ $U(1)$ symmetry
causes the rise of the mass term $m(T,\phi).$ While it exceeds the
tiny mass term $g^{1/2} m_P,$ DM and DE are dynamically coupled.  }
\label{massa}
\end{figure}

To discuss the whole dynamics in a quantitative way, let us start from
the Lagrangian
\begin{equation}
{\cal L} =  \sqrt{-g} \{ g_{\mu\nu} 
\partial_\mu \Phi^* \partial_\nu \Phi   - V(\Phi) \}~,
\end{equation}
which is $U(1)$ invariant. Let us then add to it terms of the form
(\ref{vg})
\begin{equation}
\label{due}
V_n = 4g_n {(\Phi^* \Phi)^{n+4 \over 2} \over m_P^n}
- 2g_n {(\Phi^* \Phi)^{n+3 \over 2} \over m_P^n} (\Phi+\Phi^*)~,
\end{equation}
meant to fulfill the {\it no--hair }theorem, and the terms explicitly
breaking the $U(1)$ symmetry when the chiral symmetry is broken.
Altogether $\cal L$ reads
\begin{eqnarray}
{\cal L} = \sqrt{-g} \left\{ {1 \over 2} 
g_{\mu\nu} [\partial_\mu \phi \partial_\nu \phi 
+ \phi^2  \partial_\mu \theta \partial_\nu \theta] 
- V(\phi) - \tilde m^2(T,\phi) \phi^2 (1 - \cos \theta) 
\right\} ~,
\nonumber 
\end{eqnarray}	
\begin{equation} 
\label{eq:m1}
\end{equation}
if $\phi$ and $\theta$ are explicitly used. Here $g_{\mu\nu}$ is the
metric tensor; we assume that $ds^2 = g_{\mu\nu} dx^\mu dx^\nu = a^2
(d\tau^2 - \eta_{ij}dx_idx_j)$, so that $a$ is the scale factor,
$\tau$ is the conformal time; Greek (Latin) indexes run from 0 to 3 (1
to 3); dots indicate differentiation in respect to $\tau$. The
equations of motion, for the $\phi$ and $\theta$ degrees of freedom,
read
\begin{equation}
\ddot \theta + 2\left({\dot a / a}+{\dot \phi / \phi}\right)
\dot \theta + a^2 \tilde m^2 \sin \theta = 0~,
\label{eq:m3}
\end{equation}
\begin{equation}
\ddot \phi + 2 (\dot a / a) \dot \phi + a^2 {\partial \over \partial \phi}
\left( V(\phi) + \tilde V_n \right)= \phi\, \dot \theta^2
\label{eq:m4}
\end{equation}
In general 
\begin{equation}
\label{m2t}
\tilde m^2 = m^2(T,\phi) + g_n \phi^2 (\phi/\sqrt{2}m_P)^n
\end{equation}
is made by two terms. According to \cite{KT}, at $T > \Lambda_{QCD}$
the former term exhibits a rapid rise,
\begin{equation}
m(T,\phi) \simeq 0.1 ~(\Lambda_{QCD}/T)^{3.8}~m_o(\phi)~, 
\label{e1}
\end{equation}
as $T$ approaches $\Lambda_{QCD};$ here
\begin{equation}
m_o(\phi) = q(m_q) m_\pi f_\pi /\phi
\label{e2}
\end{equation}
so that $m^2(T,\phi) \phi^2$ is $\phi$ independent.
At $T < \Lambda_{QCD},$ this term reduces to
$m_o(\phi).$ Fig.~\ref{massa} shows the rise and decline of
$m(T,\phi).$ When it prevails, DM and DE are coupled.

In the latter mass term we shall then consider just a $\tilde V_{-2}$
correction, so that
\begin{equation}
\tilde m^2 = m^2(T,\phi) + {g} m_P^2~.
\label{e3}
\end{equation}
Any value $n \neq -2$ clearly complicates the second term at the
r.h.s., yielding a $\phi$ dependent mass. In turn, such dependence
should be taken into account in the equation of motion, where the
$\partial \tilde V_n/\partial \phi$ term would become more intricate.
Only the case $n=-2$ will be treated here.

\begin{figure}[h!]
\centering
\vskip-.2truecm
\includegraphics[height=9.cm,angle=0]{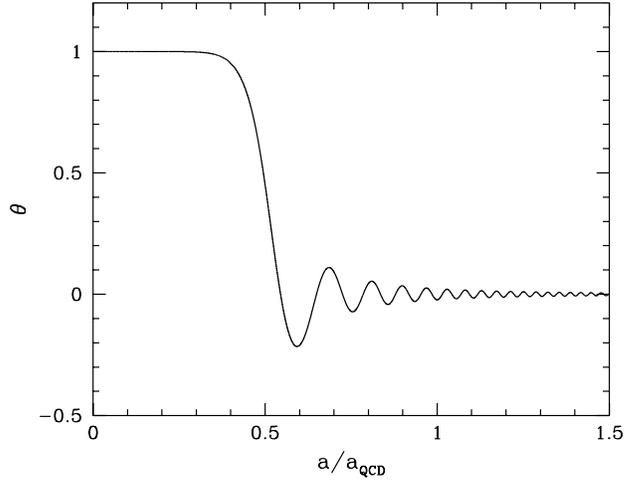}
\vskip -2.truecm
\caption{ When chiral symmetry breaks down, the PQ $U(1)$ symmetry is
also broken by the rise of the mass term $m(T,\phi).$ As a
consequence, $\theta$ becomes a significantly dynamical variable and
begins its oscillatory behavior. Here we show the results of a
numerical integration of the stages leading to $\theta$ oscillations.
}
\label{qh}
\end{figure}

In what follows, eqs.~(\ref{eq:m4}) shall be written in the form
allowed by the restriction $\theta \ll 1.$ This regime is reached soon
after quark--hadron transition, as is shown in Fig.~\ref{qh}.  In
particular, for $\theta \ll 1,$ the energy densities
$\rho_{\theta,\phi} = \rho_{\theta,\phi;kin} + \rho_{\theta, \phi;
pot}$ and the pressures $p_{\theta,\phi} = \rho_{\theta,\phi;kin} -
\rho_{\theta, \phi;pot}$ are then obtainable by combining the terms
\begin{eqnarray}	
\rho_{\theta,kin} = {\phi^2 \over 2 a^2} \dot \theta^2~,~~
\rho_{\theta,pot} =  {\tilde m^2 \over 2}
\phi^2 \theta^2~,~~ \nonumber \\ \rho_{\phi,kin} = {\dot \phi^2 \over
2 a^2} ~,~~ \rho_{\phi,pot} = V(\phi)~.~~~~~~~~
\label{enpre}
\end{eqnarray}	
When $\theta$ undergoes many (nearly) harmonic oscillations within a
Hubble time, $\langle \rho_{\theta,kin} \rangle \simeq \langle
\rho_{\theta,pot} \rangle$ and $\langle p_\theta \rangle$ vanishes
\cite{DF}.  Under such condition, using eqs.~(\ref{eq:m3}) and
(\ref{eq:m4}), it is easy to see that
\begin{equation}
\dot \rho_\theta + 3{\dot a \over a} \rho_\theta = {\dot {\tilde m}
 \over \tilde m} \rho_\theta ~,~~ \dot \rho_\phi + 3 {\dot a \over a}
 (\rho_\phi+p_\phi) = - {\dot {\tilde m} \over \tilde m} \rho_\theta ~.
\label{eq:m7}
\end{equation}
Let us notice that the r.h.s.'s of these equations will not vanish
only when $m^2(T,\phi)$ yields the dominant contribution to the
mass. This will occur about the quark--hadron transition, when it is
$\dot {\tilde m}/ \tilde m = -\dot \phi/\phi - 3.8\, \dot T/T.$ When
$T$ approches 0, however, the constant mass term dominates and the
dark component coupling fades. The exchange of energy between DM and
DE, indicated by the r.h.s.'s of the eqs.~(\ref{eq:m7}), put the
modified--DAM scheme among the set of the coupled models treated in
\cite{A}.

The coupling here depends on $\phi$ and is therefore time--dependent.
However, if we set
\begin{equation}
\label{Cop}
C(\phi) = {1 \over \phi} { m^2(T,\phi) \over \tilde m^2 }~,
\end{equation}
so that eqs.~(\ref{eq:m4}) and (\ref{eq:m7}) read, respectively,
\begin{equation}
\ddot \phi + 2 {\dot a \over a} \dot \phi + a^2 V'(\phi)  = 
C(\phi)\rho_\theta a^2
\label{eq:m4bis}
\end{equation}
\begin{equation}
\dot \rho_\theta + 3{\dot a \over a} \rho_\theta =  - C(\phi) \, \dot\phi\, 
\rho_\theta ~,~~ \dot \rho_\phi + 3 {\dot a \over a}
 (\rho_\phi+p_\phi) =  C(\phi) \, \dot\phi\, \rho_\theta
\label{eq:m7bis}
\end{equation}
we see that, when the main contribution to $\tilde m$ is given by
$m(T,\phi),$ the coefficient of $1/\phi$ in $C(\phi)$ approches unity.
In this case the time--dependence is evident.  According to
eqs.~(\ref{e1})--(\ref{e2}), however, it is $m(T,\phi) \propto
\phi^{-1}$ and the second term in eq.~(\ref{e3}) will eventually take
over.  When this occurs $ { m^2(T,\phi) /\tilde m^2 }$ becomes
negligible and the coupling between DM and DE vanishes.

In Figure \ref{massa} we describe the total mass behavior, starting
from the stage before the quark--hadron transition, in the regime when
chiral symmetry is still unbroken; then $m^2(T,\phi)$ becomes
dominant, to be overcame again by the $g\, m_P^2$ term at late times.

Altogether, DE is coupled to DM at large $z,$ but gradually decouples
at late times. The unified scheme is responsible for producing fair
amounts of DM and DE, whose origin is no longer unrelated. But one of
the extra terms aiming to reconcile the PQ approach with GR, is doomed to
hide the coupling when approaching the present epoch.

\section{Using the SUGRA potential}
If the SUGRA potential (\ref{sugra}) is used, in the radiation
dominated era, until the eve of the QH--transition, $\phi$ evolves
according to the tracker solution
\begin{equation}
\phi^{\alpha+2} = g(\alpha) \Lambda^{\alpha+4} a^2 \tau^2 ~,
\label{eq:l2}
\end{equation}
with $g(\alpha) = \alpha (\alpha+2)^2/4(\alpha+6).$ This high--$z$
tracker solution is abandoned when the coupling switches on. Then
$\dot \theta$ becomes significant so that $\phi \dot \theta^2$ exceeds
$a^2 V'$, and the field enters a different tracking regime:
\begin{equation}
\phi^{2} = {3 \over 2} \rho_{c} a^2 \tau^2 ~.
\label{eq:l3}
\end{equation}
For very small of vanishing $g$ values, this regime covers the
transition from radiation dominated to $\phi$--MD expansion, which
would actually result just in a change of the coefficient from 3/2 to 9/10.

The mass $\sqrt{g} m_P $ must be however tuned to overcome the $m(T,\phi)$
contribution to $\tilde m$ before the epoch when the cosmologically
significant mass scales enter the horizon, so to avoid the suppression
of Meszaros' effect.

\begin{figure}[h!]
\centering
\vskip-.2truecm
\includegraphics[height=9.cm,angle=0]{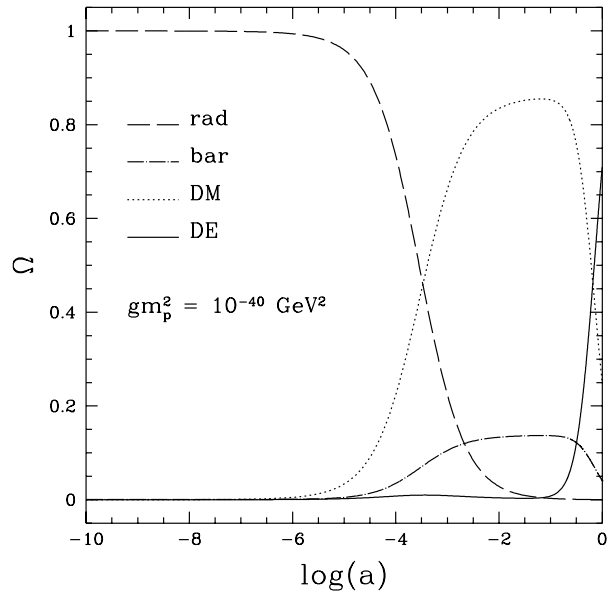}
\vskip -.5truecm
\caption{Evolution of the density parameters. Parameter behaviors are
similar to $\Lambda$CDM.}
\label{omega}
\end{figure}

\begin{figure}[h!]
\centering
\vskip-.1truecm
\includegraphics[height=9.cm,angle=0]{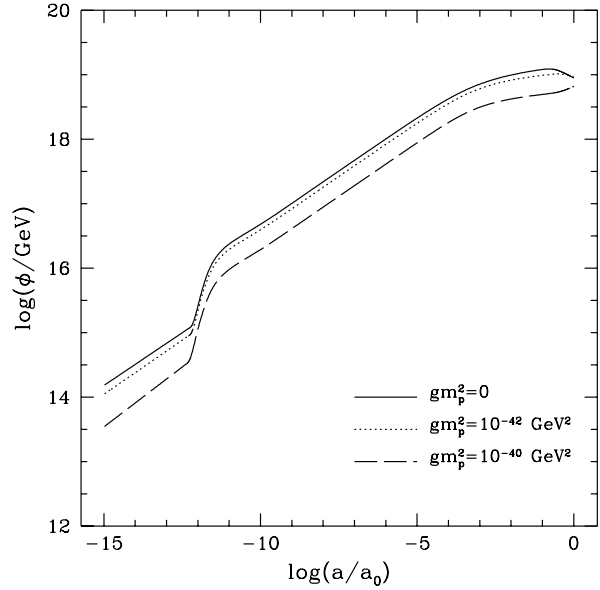}
\vskip -.5truecm
\caption{Evolution of the $\phi$ field.}
\label{phi}
\end{figure}

In Figure \ref{omega} we report the density parameters of the
different components in the various evolutionary stages, for axion
mass $g^{1/2} m_P = 10^{-20}$GeV and $\Omega_{oc}=0.25,
\Omega_{ob}=0.04,H_o=70\, $km/s/Mpc.  For the same density parameters
and $H_o$, in Figure \ref{phi} we show also the time dependence of the
$\phi$ field, for a variety of $g^{1/2} m_P$ values.


\begin{figure}[h!]
\centering
\vskip-.1truecm
\includegraphics[height=9.cm,angle=0]{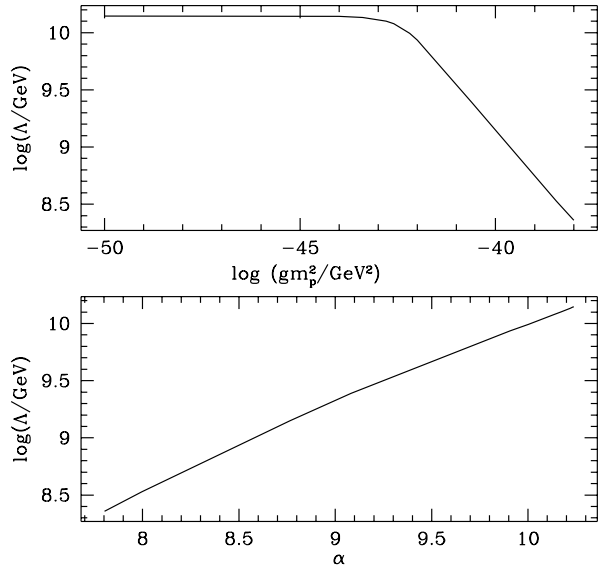}
\vskip -.5truecm
\caption{Different values of $g^{1/2} m_P $ require different values
of the energy scale $\Lambda.$ Below $\sim 10^{-22}$--$10^{-23}$GeV,
we recover the value of DAM. We report also the related values of
$\alpha.$ }
\label{la}
\end{figure}

The value of the energy scale $\Lambda,$ yielding the preferred
dark matter density parameter at $z=0,$ exhibits a dependence
on the axion mass, as is shown in Figure \ref{la}.
\begin{figure}[h!]
\centering
\vskip-.1truecm
\includegraphics[height=9.cm,angle=0]{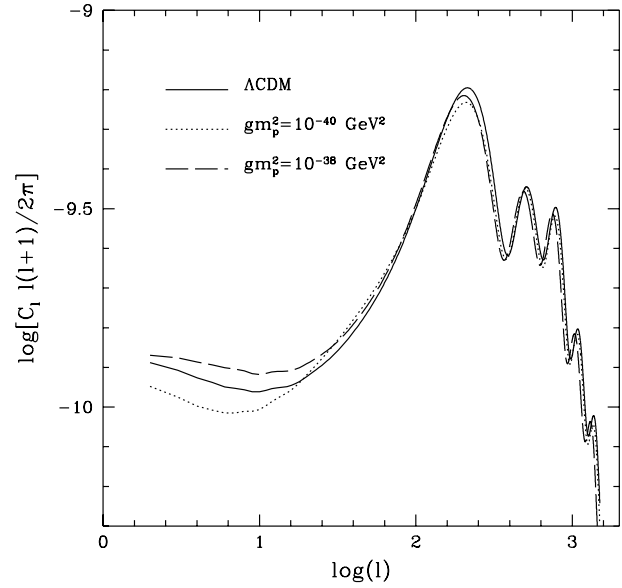}
\vskip -.5truecm
\caption{CMB anisotropy spectra for modified DAM, compared with
$\Lambda$CDM anisotropy spectra. }
\label{cl3}
\end{figure}
\begin{figure}[h!]
\centering
\vskip-.1truecm
\includegraphics[height=9.cm,angle=0]{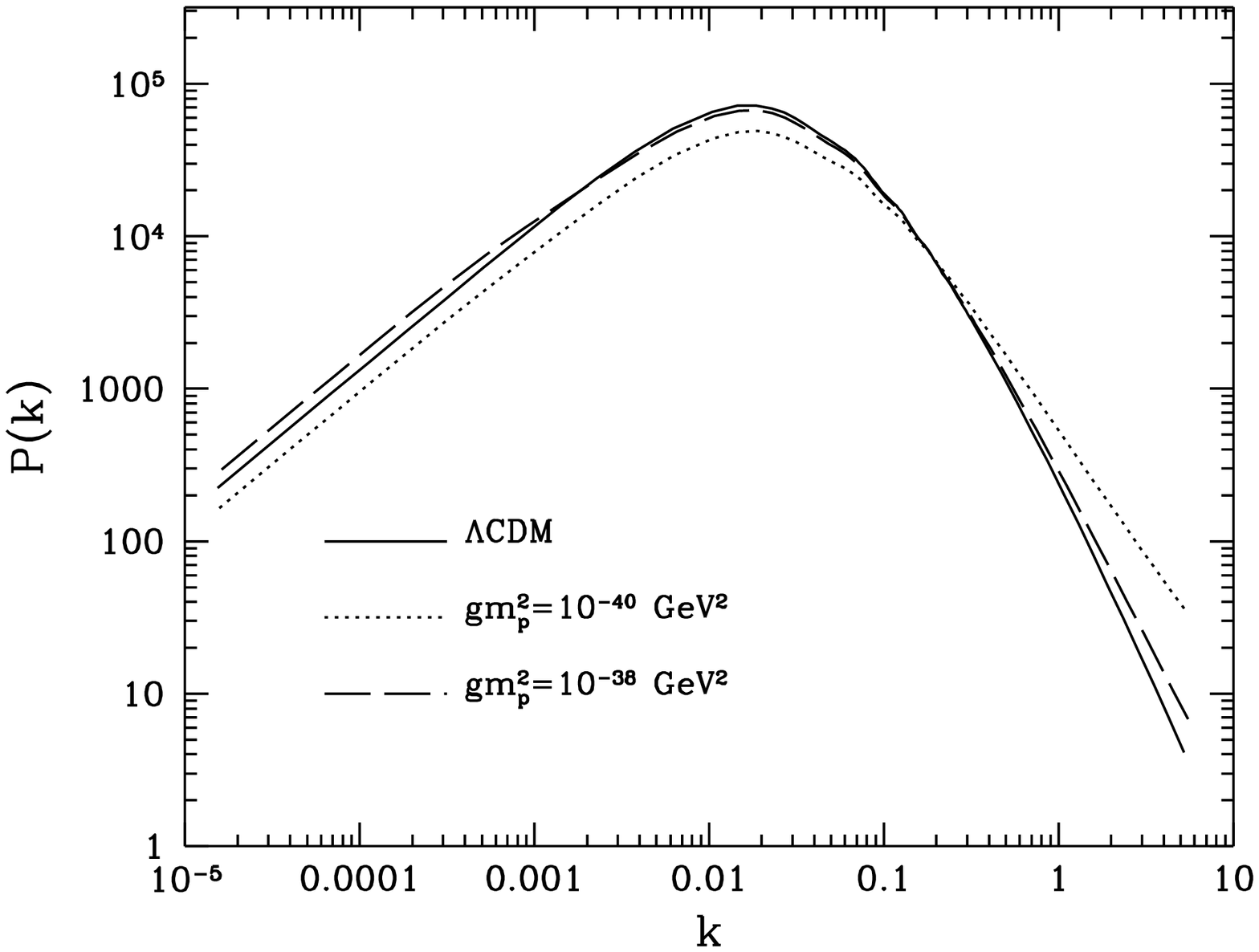}
\vskip -2.1truecm
\caption{Transfered spectra $P(k)$ for DAM, compared with
$\Lambda$CDM spectrum. }
\label{pk3}
\end{figure}

The final comparison with data, however, is to be based on CMB angular
spectra and matter fluctuation spectrum.  When considering
Fig.~\ref{cl3} it must be taken into account that all plots are
obtained with the same density parameters, $n_s$ and $H_o;$ only
normalization is slightly shifted to improve the fit. A best--fit
procedure, allowed to adapt all parameters would surely come out with
even better curves. CMB data are not a problem for DAM.

In Fig.~\ref{pk3}, then, we show the transfered spectrum for a set of
models; $n_s = 1$ and $\sigma_8 = 0.89$ are taken for all of them. $g
m_P^2$ mass values $<\sim 10^{-38}$--$10^{-39}$GeV$^2$ allow to
recover a $\Lambda$CDM--like behavior; the residual discrepancy
appearing in the Figure is mainly due to the use of dynamical DE
instad of $\Lambda$.  The Figure exhibits a progressive decrease of
the transfered spectrun steepness, when greater $g$ values are taken.

\section{Discussion and conclusions}
Reconciling PQ axion models with GR risks to spoil their elegance. PQ
model motivation is to avoid a fine tuning of the $\theta$
angle. Apparently, to correct for GR, fine tuning on ${g}^{1/2} m_P$
is needed. This problem affects both standard PQ axions and
DAM. However, although the {\it natural} mass--scale is $m_P \simeq
1.22 \times 10^{19}\, $GeV, ordinary particles are many orders of
magnitude lighter. Tuning a mass scale, therefore, is more acceptable
than tuning an angle.

Another fine--tuning problem however exists in any dynamical DE
approach. As is known, the DE field mass, obtainable from the
derivative $\partial^2 V(\phi)/ \partial \phi^2,$ in the present epoch
when $\phi \sim m_P,$ is
\begin{equation}
m_\phi^2 \simeq V(m_P)/m_P^2 \sim G\rho_{o,cr} \sim H_o^2
\end{equation}
($\rho_{o,cr} \sim V(m_P)$ is the present critical density).  Such an
extremely tiny mass, $\sim {\cal O}(10^{-42}$GeV), allows to consider
DE as a field, instead of quanta. This fine tuning is put under
further strain when potential terms $\tilde V$ (eq.~\ref{potV}) with
$p=0$ are taken, but just in association with terms with $p \neq 0$
(eq.~\ref{due}). A term with $p=0,$ with a coupling constant $g$ of
the order needed to yield axion mass, if considered autonomously,
would prevent the {\it modulus} of $\Phi$ to behave as DE.

It is true that the term we need to modify DAM, turning it into a
model quite close to $\Lambda$CDM, has the shape of terms reconciling
PQ with GR, and that a similar tuning is however necessary also to
this aim (instead of a term with $p=0$ one could then tune a
``cosmological constant'' term). But here
we need a $\tilde V_{n}$ potential with $n=-2,$ while fulfilling
the {\it no--hair} theorem requires $n>0.$ (It is however fair to add
that, in the case of large wormhole effects, the need of a $V_{-2}$
potential has also been discussed \cite{nohair}.)

It might then well be that quantum gravity does not prescribe a single
$\tilde V_n$ correction, but a combination of them. For instance,
instead of a power of $\phi/\sqrt{2} m_P$ the right potential could
naturally include a polynomial. Then, while making PQ approach
coherent with GR, the correction would also include terms explaining
why DM and DE, after a period when they interact, gradually
re--decouple, while they are driven to have similar densities in the
present epoch.

However, once discrepancies between DAM and standard dynamical DE
models are fully avoided, the possibility to falsify modified DAM
might seem limited to its particle aspects. This is only
partially true: in fact modified DAM, with a SUGRA potential, makes a
prediction on the energy range for the scale $\Lambda$ (and/or the
exponent $\alpha$) in the potential. Available data do not provide
very stringent constraints on the energy scale $\Lambda$ and the DAM
value is still consistent with them. More precise CMB data, however,
may be soon available and the energy scale $\Lambda$ will be more
stringently constrained.

But the model goes farther, predicting a relation between the precise
$\Lambda$ value, in the above scale range, and $\Omega_{oc}.$ Testing
this prediction requires still higher precision cosmological data,
which could be achievable by next generation experiments.  Meanwhile,
however, if consistency is confirmed, a precise $\Lambda$ value can be
predicted from~$\Omega_{oc}.$

\begin{ack}
Luca Amendola and Loris Colombo are gratefully
thanked for their comments on this work.
\end{ack}

{}

\end{document}